\documentclass[12pt%
 ,draft%
 ]{article}

\usepackage{amsfonts}
\usepackage{amssymb}
\usepackage{amsmath}
\usepackage{latexsym}
\usepackage{amsthm}
\usepackage{soul}

\newtheorem{theorem}{Theorem}[section]
\newtheorem{lem}[theorem]{Lemma}

\makeatletter

\def\@maketitle{%
  \newpage
  \null
  \vspace*{-2cm}%
  \begin{center}%
    {\Large\bf \@title \par}%
    \vskip 1.5em%
    {\normalsize
      \lineskip .5em%
      \begin{tabular}[t]{c}%
        \@author
      \end{tabular}\par}%
    \vskip .5em%
    \vskip 1em%
    {\large \@date}%
  \end{center}%
  \par
  \vskip 1.5em}

\def\keywords#1{\par\vskip 1.5ex%
    {\footnotesize{\noindent{\it Keywords}\/:\ #1\par}}%
    }

\def\subjclass#1{\par\vskip1mm%
    {\footnotesize{\noindent{\it Subjclass}\/:\ #1\par}}%
    }

\def\section{\@startsection {section}{1}{0pt}%
    {-4.5ex \@plus -.2ex \@minus -.1ex}%
    {2.3ex \@plus .2ex \@minus .1ex}%
    {\large\bfseries\boldmath}}

\renewenvironment{proof}[1][\proofname]{\par
  \pushQED{~\qed}%
  \normalfont \topsep3\p@\@plus3\p@\relax
  \trivlist
  \item[\hskip\labelsep
        \bfseries\boldmath
    #1\@addpunct{.}] \ignorespaces
}{%
  \popQED\endtrivlist\@endpefalse
}

\makeatother

\title{A two-point boundary value problem on a~Lorentz manifold
arising in A.~Poltorak's concept of reference frame}

\author{%
 {\large Yu.E.~Gliklikh}\\[2mm]
 Mathematics Faculty\\
 Vrornezh State University\\
 Universitetskaya pl., 1\\
 394006 Voronezh,
 Russia\\
 {\it yeg@math.vsu.ru}
 \and
 {\large P.S.~Zykov}\\[2mm]
 Physics and
 Mathematics Faculty\\
 Kursk State University\\
 ul. Radishcheva, 33\\
 305416 Kursk, Russia\\
 {\it petya39b@mail.ru}%
 }

\date{}

\begin{document}

\maketitle

\begin{abstract}

\noindent In A.~Poltorak's concept, the reference frame in General
Relativity is a certain manifold equipped with a connection. The
question under consideration here is whether it is possible to join
two events in the space-time by a time-like geodesic if they are
joined by a geodesic of the reference frame connection that has a
time-like initial vector. This question is interpreted as whether an
event belongs to the proper future of another event in the
space-time in case it is so in the reference frame. For reference
frames of two special types some geometric conditions are found
under which the answer is positive.



\end{abstract}

\def\thefootnote{*}
\footnotetext{The research is supported in part by RFBR Grant No.
04-01-00081.}%

\section{Introduction}
\hspace{2mm}Let ${\sf M}$ be a $4$-dimensional Lorentz manifold
(space-time of General Relati\-vity, for principal notions of GR,
see \cite{MTW,121}). For definiteness sake, hereafter we use the
Lorentz metric with signature $(-,+,+,+)$.

In \cite{Poltorak1,Poltorak2} A.~Poltorak suggested a concept in
which a reference frame in GR is defined as a certain smooth
manifold with a connection. In the most simple cases, this is the
Minkowski space with its natural flat connection but, in more
complicated cases, some more general manifolds and connections may
also appear.

In the reference frame, the Gravitation field is described as a
$(1,2)$-tensor $G$ that on any couple of vector fields $X$ and $Y$
takes the value
$$G(X,Y)=\nabla_XY -\bar{\nabla}_XY,
$$
where $\bar\nabla$ is the covariant derivative of the Levi--Civita
connection of the Lorentz metric while $\nabla$ is the covariant
derivative of the connection in the reference frame. Denote by
$\frac{D}{d\tau}$ the covariant derivative of the connection in the
reference frame along a given world line with respect to a certain
parameter $\tau$. Then the geodesic $m(\tau)$ of the Levi--Civita
connection in ${\sf M}$ (a world line in the absence of force fields
except the gravitation) is described in the reference frame by the
equation (here $G_m(X, Y)$ is the value of $G(X, Y)$ at point $m$)
\begin{equation}
\label{eq1}
\frac{D}{d\tau}m'(\tau)=G_{m(\tau)}(m'(\tau),m'(\tau)).
\end{equation}
Notice that the right-hand side of \eqref{eq1} is quadratic in
velocities $m'(\tau)$.

We refer the reader to \cite{Poltorak1,Poltorak2} for more details
about the Poltorak's concept and for physical interpretation of the
covariant derivative of connection in the reference frame, of the
tensor $G$ and several other objects associated with it. The
subsequent development of this idea can be found in
\cite{Poltorak3}.

We suggest a version of the concept where the manifold of reference
frame is the tangent space ${\sf T}_m {\sf M}$ at an event $m\in
{\sf M}$ and a Lorentz-orthonormal basis $e_\alpha$, where
$\alpha=0,1,2,3$, is specified (the time-like vector $e_0$ is the
observer's $4$-velocity). We suppose that this reference frame is
valid in a neighborhood $\cal O$ of the origin in ${\sf T}_m{\sf
M}$, which is identified with a neighborhood $\cal U$ of the event
$m$ by the exponential map of the Levi--Civita connection of the
Lorentz metric (the normal chart).

We deal with two choices of connection on the manifold ${\sf T}_m
{\sf M}$. In the first one, we consider on ${\sf T}_m {\sf M}$ its
natural flat connection of Minkowski space (the main case taken into
account by A.~Poltorak). In the second case, we involve a Riemannian
connection of a certain (positive definite) Riemannian metric on
${\sf T}_m {\sf M}$. This case is motivated by a natural development
of the idea yielding Euclidean models in the Quantum Field Theory.
Observe that the above-mentioned Riemannian connection may not be
the Levi--Civita one, and a non-zero torsion connection compatible
with the metric is also allowed. In principle, this allows us to
consider electromagnetic interactions as well.

For two reference frames mentioned above, we investigate the
question of whether it is possible to connect two events $m_0$ and
$m_1$ in ${\sf M}$ by a time-like geodesic if they are connected in
the reference frame by a geodesic of the corresponding connection
whose initial vector is time-like, i.e., lies inside the light cone
in the space ${\sf T}_{m_0}{\sf M}$. This question can be
interpreted as follows: does the event $m_1$ belong to the proper
future of the event $m_0$ if this is true in the reference frame? We
find geometric conditions under which the answer to this question is
positive.

In what follows, we use the following technical statement.

\begin{lem}\label{1}
Specify arbitrary positive real numbers $\varepsilon$, ${T}$ and
$C$. Let a number $b$ be such that
$0<b<\frac{\varepsilon}{(\varepsilon + C)^2}$. Then there exists a
sufficiently  small positive real number  $\varphi$ such that
\begin{equation}
\label{eq*}
 b((\varepsilon { T}^{-1}-\varphi)+ C{ T}^{-1})^2<\varepsilon
 { T}^{-2}-\varphi { T}^{-1}.
\end{equation}
\end{lem}

\begin{proof} For $b$ satisfying the hypothesis of the Lemma,
we have
$$b(\varepsilon { T}^{-1}+ C{ T}^{-1})^2<\varepsilon {
T}^{-2}. $$ The continuity obviously implies that there exists a
sufficiently small real number $\varphi >0$ such that
$(\varepsilon { T}^{-1}-\varphi)>0$ and \eqref{eq*} holds.
\end{proof}

\section{The reference frame with flat connection}

In this section we investigate the reference frame 
of the first type, mentioned above, i.e., the manifold of reference
frame is ${\sf T}_m{\sf M}$ with a certain basis $e_\alpha$, where
$\alpha=0,1,2,3$,  such that the time-like vector $e_0$ is the
$4$-velocity of a certain observer, and the connection in ${\sf
T}_m{\sf M}$ is the flat connection of the Minkowski space.

In this case, it is convenient to consider $\cal O$ as a domain in a
linear space, on which there are given the Lorentz metric, the
tensor $G$ and other objects described in Introduction. It is also
convenient to use the usual facts of Linear Algebra. In particular,
the tangent space ${\sf T}_{\bar m}{\sf M}$ at $\bar m\in {\cal O}$
can be identified with ${\sf T}_m{\sf M}$ by a translation and for
any $\bar m \in {\cal O}$, we may consider the light cone in ${\sf
T}_{\bar m}{\sf M}$ (generated by Lorentz metric tensor at $\bar m$)
as lying in ${\sf T}_m{\sf M}$ but depending on $\bar m$.

Geodesic lines in ${\sf T}_m {\sf M}$ with respect to the flat
connection are straight lines. Thus, in this reference frame, the
question under consideration takes the following form: \so{is it
possible to connect the events $m_0$ and $m_1$ on ${\sf M}$ by a
time-like geodesic if they are connected by the straight line
$a(\tau)$ in $\cal O$ so that $a(0)=m_0$, $a({ T})=m_1$ and that
lies inside the light cone in ${\sf T}_{m_0}{\sf M}$?} Here $\tau$
is a certain parameter that can be, say, the proper time on ${\sf
M}$ or the natural parameter in the reference frame, etc. Notice
that in this case the fact that the straight line $a(\tau)$
belongs to the light cone in ${\sf T}_{m_0}{\sf M}$, is equivalent
to the fact that the vector of derivative
$a'(0)=\frac{d}{d\tau}a(\tau)_{|\tau=0}$ lies inside that cone, as
it is postulated in the problem under consideration.

Since the covariant derivative with respect to the flat connection
coincides in this case with the ordinary derivative in ${\sf
T}_m{\sf M}$, equation \eqref{eq1} takes the form
\begin{equation}
\label{eq2}
\frac{d}{d\tau}{m'}(\tau) =G_m(m',m'). 
\end{equation}
Thus the main problem is reduced to the two-point boundary value
problem for \eqref{eq2}. Since the right-hand side of \eqref{eq2}
has quadratic growth in velocities, for some couples of points the
two-point problem may have no solutions.

Recall that the tangent space ${\sf T}_m {\sf M}$ to the Lorentz
manifold ${\sf M}$ has the natural structure of a Minkowski space
whose scalar product is the metric tensor of ${\sf M}$ at the event
$m$. Since the specified basis $e_\alpha$, where $\alpha=0,1,2,3$,
is Lorentz-orthonormal, their Minkowski scalar product of
$X=X^\alpha e_\alpha$ and $Y=Y^\alpha e_\alpha$  has the form
\[
X\cdot Y=-X^0Y^0+X_iY^i,
\]
where $X_i=X^i$ for $i=1,2,3$ (we use
the Einstein's summation convention). Introduce the Euclidean scalar
product in ${\sf T}_m{\sf M}$ by changing the sign of the time-like
summand, i.e., by setting
\[
(X,Y)=X^0Y^0+X_iY^i.
\]
Hereafter in this section all norms and distances are determined
with respect to the latter scalar product.

By a linear change of time introduce a parameter $s$ along
$a(\cdot)$ such that for the line $\tilde a(s)$ obtained from
$a(\tau)$, we get $\tilde a(0)=m_0$ and $\tilde a(1)=m_1$. Consider
the Banach space $C^0([0,1],{\sf T}_{m}{\sf M})$ of continuous
curves in ${\sf T}_{m}{\sf M}$ with the usual supremum norm.

\begin{lem}\label{11}
There exists a sufficiently small real number $\varepsilon>0$ such
that, for any curve $\tilde v(s)$ from the ball
$U_{\varepsilon}\subset C^0([0,1],{\sf T}_{m}{\sf M})$ of radius
$\varepsilon$ centered at the origin, there exists a vector $\tilde
C_{\tilde v}\in {\sf T}_m{\sf M}$ belonging to a certain bounded
neighborhood of the vector $\tilde a'(0)=\frac{d}{ds}{\tilde
a(s)}_{|s=0}$ and such that $\tilde C_{\tilde v}$ lies inside the
light cone of the space ${\sf T}_{m_0}{\sf M}$ and the curve
$m_0+\int_{0}^{s} (\tilde v(t) + \tilde C_{\tilde v})dt$ takes the
value $m_1$ at $s=1$. The vector $\tilde C_{\tilde v}$ continuously
depends on $\tilde v(\cdot)$  and $\|\tilde C_{\tilde v}\|<C$ for
any curve $\tilde v \in U_\varepsilon$ for some $C>0$.
\end{lem}

\begin{proof} By  explicit  integration
one can easily prove that $C_{\tilde v}$ such that $m_0+\int_{0}^{1}
(\tilde v(t) + \tilde C_{\tilde v})dt=m_1$ does exist and that it is
continuous in $\tilde v$. Then by continuity, from the fact that the
vector $\tilde a'(0)$ lies inside the light cone of the space ${\sf
T}_{m_0}{\sf M}$, it follows that for a perturbation $\tilde
v(\cdot)$ sufficiently small with respect to the norm, the vector
$\tilde C_{\tilde v}$ also lies inside the same light cone. Take for
$C$ the upper bound of the set of norms of vectors $\tilde C_{\tilde
v}$ from the above-mentioned bounded neighborhood of
$\frac{d}{ds}{\tilde a(s)}_{|s=0}$.~\end{proof}

Notice that $C$ is an estimate of Euclidean distance between
$m_0$ and $m_1$.

Turn back to the parametrization of the line $a(\cdot)$ by the
parameter $\tau$. Consider the Banach space $C^0([0,{ T}],{\sf
T}_{m}{\sf M})$.

\begin{lem}\label{12}
Let a real number $k>0$ be such that ${ T}^{-1}\varepsilon
>k$, where $\varepsilon$ is from lemma \ref{11}. Then for any curve $v(t)$
from the ball $U_{k}\subset C^0([0,{ T}],{\sf T}_{m}{\sf M})$ of
radius $k$ centered at the origin, there exists a vector $C_{v}\in
{\sf T}_m{\sf M}$ from a bounded neighborhood of the vector
$a'(0)=\frac{d}{d\tau}{a(\tau)}_{|\tau=0}$ such that the vector
$C_{v}$ lies inside the light cone of the space ${\sf T}_{m_0}{\sf
M}$ and the curve $m_0+\int_{0}^{\tau} (v(t) + C_{v})dt$  takes the
value $m_1$ at $\tau={ T}$. The vector $C_{v}$ is continuous in
$v(\cdot)$.
\end{lem}

\begin{proof} Changing the time along $a(\tau)$ construct the
straight line $\tilde a(s)=a({ T}s)$ that meets the conditions
$\tilde a(0)=m_0$ and $\tilde a(1)=m_1$ as in Lemma \ref{11}. For
any curve $v(\cdot)\in U_k\subset C^0([0,{ T}],{\sf T}_{m}{\sf M})$,
construct the curve $\tilde v(s)={ T}v({ T}s)$ that lies in
$U_{\varepsilon}\subset C^0([0,1],{\sf T}_{m}{\sf M})$, i.e., such
that satisfies Lemma \ref{11}. In particular, for this curve, there
exists a vector $\tilde C_{\tilde v}$ such that $\|\tilde C_{\tilde
v}\|<C$ from Lemma \ref{11}. By explicit calculations one can easily
derive that
$$m_0+\int_0^1(\tilde v(s)+\tilde C_{\tilde
v})ds=m_0\int_0^{ T}(v(t)+C_{v})dt=m_1, $$ where $C_v={
T}^{-1}\tilde C_{\tilde v}$.
\end{proof}

Notice that by construction $\|C_v\|<{ T}^{-1}C$ for any $v\in U_k$.

For the tensor $G$, introduced above, define its norm $\|G_m\|$ by
the standard formula $\|G_m\|= \sup\limits_{X\in {\sf T}_m{\sf
M},\|X\|\le 1}\|G_m(X,X)\|$. The definition immediately implies the
estimate
\begin{equation}
\label{eq3}
\|G_m(X,X)\|\le \|G_m\|\|X\|^2 \; \text{ for
 any } X\in {\sf T}_m{\sf M}.
\end{equation}

\begin{theorem}
Let $m_0$ and $m_1$ be connected in $\cal O$ by a straight line
$a(\tau)$ that lies inside the light cone of the space ${\sf
T}_{m_0}{\sf M}$ and satisfies the conditions $a(0)=m_0$ and $a({
T})=m_1$. Let $m_0$ and $m_1$ belong  to a ball $V\subset {\sf
T}_{m}{\sf M}$ such that for any $\hat m\in V$ the inequality
$\|G_{\hat m}\|<\frac{\varepsilon}{(\varepsilon+C)^2}$ holds, where
$\varepsilon$ and $C$ are from Lemma \ref{11}. Then on ${\sf M}$
there exists a time-like geodesic $m_0(\tau)$ of the Levi--Civita
connection of the Lorentz metric such that $m_0(0)=m_0$ and $m_0({
T})=m_1$.
\end{theorem}

\begin{proof} Consider the ball $U_K\subset C^0([0,{ T}],{\sf
T}_{m}{\sf M})$ of radius $K={ T}^{-1} \varepsilon - \varphi$
centered at the origin, where $\varphi$ is from Lemma \ref{1}. Since
$K <{ T}^{-1}\varepsilon$, the assertion of Lemma \ref{12} is true
for $U_K$ and the following completely continuous operator
$$
Bv= \int_{0}^{\tau}G_{m_0+\int_{0}^{t}(v(s)+C_{v})ds}(v(t)+C_{v},
v(t)+C_{v})dt
$$
is well posed on this ball. Let us show that this operator has a
fixed point in $U_K$. Recall that, for any curve $v\in U_K$, its
$C^0$-norm is not greater than $K={ T}^{-1} \varepsilon - \varphi$
and that by Lemma \ref{12} $\|C_v\|<{ T}^{-1}C$. Then  the
hypothesis of the theorem, eq. \eqref{eq3} and Lemmas \ref{1},
\ref{11} and \ref{12} imply that
\[
\begin{split}
 &\|G_{m_0+\int_{0}^{t}(v(s)+C_{v})ds} \left( (v(t) + C_{v}),
 (v(t)+C_v) \right)\|\le\\
 &\|G_{m_0+\int_{0}^{t}(v(s)+C_{v})ds}\| ((\varepsilon{T}^{-1}-\varphi)+ C{ T}^{-1})^2<
 ({ T}^{-2}\varepsilon-{T}^{-1}\varphi).
\end{split}
\]
 From the last inequality we obtain
\[
\bigg\|\int_{0}^{\tau}G_{m_0+\int_{0}^{t}(v(s)+C_{v})ds}((v(t)+C_{v}),
(v(t)+C_{v}))dt\bigg\| < ({ T}^{-1}\varepsilon - \varphi)=K.
\]
This means that the operator ${\cal B}$ sends the ball $U_K$ into
itself and so, by Schauder's principle, it has a fixed point
$v_{0}(t)$ in this ball. It is easy to see that $m_0(\tau)
=m_0+\int_0^\tau (v_{0}(t)+C_{v_{0}})dt$ is a solution of
differential equation \eqref{eq2} such that $m_0(0)=m_0$ and
$m_0({ T})=m_1$. Notice that by the construction $m_0(\tau)$ is a
geodesic of the Levi--Civita connection of the Lorentz metric on
${\sf M}$. The equality ${\cal B}v_{0}=v_{0}$ and the definition
of $\cal B$ implies that $v_0(0)=0$, and hence
$\frac{d}{d\tau}m_0(\tau)_{|\tau=0}=C_{v_0}$, where by Lemma
\ref{12} the vector $C_{v_0}$ lies in the light cone of the space
${\sf T}_{m_0}{\sf M}$, i.e., its Lorentz scalar square is
negative. Since this scalar square of the vector of derivative is
constant along the geodesic of Levi--Civita connection of the
Lorentz metric, this geodesic is time-like.
\end{proof}

\section{The reference frame with Riemannian connection}

In this section we investigate the reference frame at the event
$m\in {\sf M}$ of the second type, mentioned in Introduction.
Namely, the manifold here is absolutely the same as in the
previous section: ${\sf T}_m{\sf M}$ with a specified orthonormal
basis, while the connection may not be flat but it is supposed to
be compatible with a certain (positive definite) Riemannian metric
on ${\sf T}_m{\sf M}$ (see Introduction). Recall that we do not
assume this connection  be torsion-less.


Here, the question of existence of a geodesic of Levi--Civita
connection on ${\sf M}$, that we are looking for, is reduced to
the solvability of the two-point boundary value problem for
equation \eqref{eq1} in the reference frame.

An important difference of this case from the case of flat
connection is the fact that, for non-flat connections, the
conjugate points may exist. There are examples (see
\cite{yeg,yeg2}) showing that, for a couple of points conjugate
along all geodesics joining them, the boundary value problem for a
second order differential equation may have no solutions at all,
even if its right-hand side is smooth and  bounded. Besides, as it
is mentioned in the previous section, the two point boundary value
problem for \eqref{eq1} may be not solvable since the right-hand
side of \eqref{eq1} has quadratic growth in velocities. We suppose
from the very beginning that $m_0$ and $m_1$ are connected by a
geodesic in the reference frame, along which they are not
conjugate. In this situation, we find conditions under which the
problem for \eqref{eq1} is solvable.

The case of non-flat connection requires more complicated
machinery than the previous one. In particular, we replace
ordinary integral operators (used in the previous section) by {\it
integral operators with parallel translation} introduced by
Yu.~Gliklikh (see, e.g., \cite{yeg,yeg2}).

Everywhere in this section the norms in the tangent spaces and the
distances on manifolds are induced by the above positive definite
Riemannian metric.

First, we describe some general constructions. Let $\cal M$ be a
complete Rie\-mannian manifold, on which a certain Riemannian
connection is fixed. Consider ${p}_0 \in {\cal M}$, $I=[0,l] \subset
R$ and let $v:I \to {\sf T}_{p_{0}} {\cal M}$ be a continuous curve.
Applying a construction of Cartan's development type, one can show
(see, e.g., \cite{yeg,yeg2}) that there exists a unique $C^1$-curve
$p:I \to {\cal M}$ such that $p(0)=p_0$ and the vector $p' (t)$ is
parallel along $p(\cdot)$ to the vector $v(t) \in {\sf
T}_{p_{0}}{\cal M}$ for any $t \in I$.

Let $p(t):=Sv(t)$ be the curve constructed above from the curve
$v(t)$. Thus the continuous operator $S:C^{0}(I,{\sf T}_{p_0} {\cal
M}) \to C^{1}(I,{\cal M})$ that sends the Banach space $C^{0}(I,{\sf
T}_{p_0} {\cal M})$ of continuous curves in ${\sf T}_{p_0}{\cal M}$
to the Banach manifold $C^{1}(I,{\cal M})$ of $C^1$-curves in ${\cal
M}$ (the domain of all curves is $I$)  is well-posed.

Notice that for a constant curve $v(t)\equiv X\in {\sf
T}_{p_0}{\cal M}$ we get by construction that $Sv(t)=\exp X$,
where $\exp$ is the exponential map of the given connection.

Instead of ${\cal M}$ we can consider the neighborhood $\cal O$ in
${\sf T}_m{\sf M}$ described in Introduction. Without loss of
generality we may assume that the Riemannian metric on $\cal O$ is a
restriction of a certain complete Riemannian metric on ${\sf
T}_{m}{\sf M}$. Indeed, take a relatively compact domain ${\cal
O}_1\subset {\cal O}$ with smooth boundary such that ${\cal O}_1$
contains the points $0\in {\sf T}_{m_0}{\sf M}$, $m_0$ and $m_1$ as
well as the geodesic $\gamma(t)$, where $t\in [0,1]$, joining $m_0$
and $m_1$ (if $\cal O$ is relatively compact one can take it
for${\cal O}_1$). Then it is possible to change the Riemannian
metric outside ${\cal O}_1$ so that it becomes complete on ${\sf
T}_{m}{\sf M}$, and to use ${\cal O}_1$ instead of $\cal O$. Thus,
the operator $S$ is well-posed in this situation.

Let the points $m_0, m_1\in {\cal O}$ be connected in $\cal O$ by a
geodesic $\gamma(t)$ of the Riemanninan connection so that
$\gamma(0)=m_0$ and $\gamma(1)=m_1$. In particular, we get
$m_1=\exp(\frac{d}{dt}\gamma(t)_{|t=0})=S(\frac{d}{dt}\gamma(t)_{|t=0})$,
where $\exp$ is the exponential map of the Riemannian connection.
Let $m_0$ and $m_1$ be not conjugate along $\gamma(\cdot)$ and  the
vector $\frac{d}{dt}\gamma(t)_{|t=0}$ lie inside the light cone of
${\sf T}_{m_0}{\sf M}$.

Hereafter we denote by $U_k$ the ball of radius $k$ centered at the
origin in a certain Banach space of continuous maps.

\begin{lem}\label{2} Let $\gamma(t)$ be a geodesic of the connection
in the reference frame such that $\gamma(0)=m_0$ and
$\gamma(1)=m_1$. Let also $m_0$ and $m_1$ be non-conjugate along
$\gamma(\cdot)$. Then there exists a number $\varepsilon>0$ and a
bounded neighborhood $V$ of the vector
$\frac{d}{dt}{\gamma}(t)_{|t=0}$ in ${\sf T}_{m_0} {\cal O}$ such
that, for any curve $\tilde {u}(t) \in U_{\varepsilon} \subset C^0
([0,1],{\sf T}_{m_0} {\cal O})$, the neighborhood $V$ contains a
unique vector $\tilde C_{\tilde {u}}$, depending continuously on
$\tilde {u}$, such that $S(\tilde {u}+\tilde C_{\hat{u}})(1)=m_{1}$.
\end{lem}

Denote by $C$ an upper bound of the norms of vectors $\hat C_{\tilde
{u}}$ from Lemma~\ref{2}.

\begin{lem}\label{3}
In conditions and notations of Lemma \ref{2}, let the numbers $k>0$
and ${ T}>0$ satisfy the inequality ${ T}^{-1} \varepsilon
> k$. Then, for any curve $u(t) \in U_k\subset C^0
([0,{ T}],{\sf T}_{m_0} {\cal O})$ in a certain bounded neighborhood
of the vector ${ T}^{-1}\frac{d}{dt}{\gamma}(t)_{|t=0}$ in ${\sf
T}_{m_0} {\cal O}$, there exists a unique vector $C_u$, depending
continuously on $u$, such that $S(u+C_u)({ T})=m_1$.
\end{lem}

The proofs of Lemmas \ref{2} and \ref{3} are quite analogous to that
of Theorem 3.3 in \cite{yeg} (see also proofs of Lemmas 1 and 2 in
\cite{zyk}). It should be pointed out that, as in Lemma \ref{12}, we
have $C_v={ T}^{-1}C_{\tilde v}$, where $\tilde v(s)={ T}v({ T}s)\in
U_\varepsilon\subset C^0 ([0,1],{\sf T}_{m_0} {\cal O})$. Thus, for
$u\in U_k$ from Lemma \ref{3}, the inequality $\|C_u\|<{ T}^{-1}C$
holds, where $C$ is introduced after Lemma \ref{2}.

\begin{lem}\label{4} In the conditions and notations of Lemmas
\ref{2} and \ref{3}, the number $\varepsilon$ can be chosen so that
for the curve $\tilde u(\cdot)\in U_\varepsilon\subset C^0
([0,1],{\sf T}_{m_0} {\cal O})$ the vector $\tilde C_{\tilde u}$
lies inside the light cone of the space ${\sf T}_{m_0}{\sf M}$ and,
for the curve $u\in U_k\subset C^0 ([0,{ T}],{\sf T}_{m_0} {\cal
O})$, the vector $C_u$ also lies inside the light cone of the space
${\sf T}_{m_0}{\sf M}$.
\end{lem}

\begin{proof} The fact that,  for sufficiently small $\varepsilon>0$,
the vector $C_{\tilde u}$ belongs to the interior of the light cone,
is derived from continuity consideration as in Lemma \ref{11}. For
$C_u$, this statement follows from the fact that $C_v={
T}^{-1}C_{\tilde v}$, where $\tilde v(s)={ T}v({ T}s)\in
U_\varepsilon\subset C^0 ([0,1],{\sf T}_{m_0} {\cal O})$ (see
above).
\end{proof}

Hereafter we choose $\varepsilon$ satisfying the hypotheses of
Lemmas \ref{2} and \ref{4}.

Let $\gamma(t)$ be a $C^1$-curve given for $t\in [0,{ T}]$, let
$X(t,m)$ be a vector field on ${\cal O}$. Denote by $\Gamma
X(t,\gamma(t))$ the curve in ${\sf T}_{\gamma(0)}{\cal O}$ obtained
by parallel translation of vectors $X(t,\gamma(t))$ along
$\gamma(\cdot)$ at the point $\gamma(0)$ with respect to the
connection of the reference frame.

In complete analogy with the previous section, we introduce the
norm $\|G_m\|$ by means of the norms of vectors with respect to
the Riemannian metric on $\cal O$ as it was mentioned above.
Obviously, eq. \eqref{eq2} is valid for $\|G_m\|$, as in the
previous section.

With the help of $S$ and $\Gamma$ we construct the integral operator
$B: U_k \rightarrow C^0([0,{ T}], {\sf T}_{m_0}{\sf M})$ of the
following form:
\begin{equation}
\label{eq4}
 Bv= \int_{0}^{\tau} \Gamma G_{S(v(t) + C_{v})}
 \left(\frac{d}{dt}S(v(t)+C_v) , \frac{d}{dt}S(v(t)+C_v) \right)dt,
\end{equation}
where $k$ and ${ T}$ satisfy the conditions of Lemma \ref{3}. One
can easily see that the operator $B$ is completely continuious.

\begin{theorem}
Let $m_0,m_1\in {\cal O}$ and let there exist a geodesic $\gamma
(\tau)$ of the connection of the reference frame such that $\gamma
(0)=m_0$, $\gamma({ T})=m_1$, $m_0$ and $m_1$ are not conjugate
along $\gamma (\cdot)$ and the vector $\frac{d}{dt}\gamma(t)_{|t=0}$
lies inside the light cone of the space ${\sf T}_{m_0}{\sf M}$. If
$m_0$ and $m_1$ belong to the ball $V\subset{\cal O}$, such that at
any $m\in V$ the inequality
$\|G_m\|<\frac{\varepsilon}{(\varepsilon+C)^2}$ holds, where
$\varepsilon$ and $C$ are from Lemmas \ref{2} and \ref{4}, then
there exists a time-like geodesic $m_0(\tau)$ of the Levi--Civita
connection of Lorentz metric on ${\sf M}$ such that $m_0(0)=m_0$ and
$m_0({ T})=m_1$.
\end{theorem}

\begin{proof} Let $k:={ T}^{-1} \varepsilon -
\varphi$, where $\varphi$ is from Lemma \ref{1}. For this $k$, the
hypothesis of Lemma \ref{3} is satisfied. Hence, on the ball $U_k
\subset C^0([0,{\sf T}_1],{\sf T}_{m_0}{\sf M})$, the operator
\eqref{eq4} is well-posed. Recall that, for the curve $v(\cdot)\in
U_k$, its $C^0$-norm is not greater than $k$, that $\|C_v\|<{
T}^{-1}C$ and that the parallel translation with respect to the
Riemannian connection preserves the norms of vectors. Then taking
into account the definition of operator $S$ we see that
$$\|\frac{d}{d\tau}S(v(\tau)+C_v)\|< ({ T}^{-1} \varepsilon -
\varphi)+{ T}^{-1}C. $$ Hence,  eq. \eqref{eq2}, the hypothesis of
the theorem and  Lemma \ref{1} imply that
\begin{equation}
\begin{split}
 &\Big\|G_{ S(v(\tau) + C_{v})} \Big(\frac{d}{d\tau}S(v(\tau)+C_v),
 \frac{d}{d\tau}S(v(\tau)+C_v) \Big)\Big\| \le \\
 &\|G_{ S(v(\tau) + C_{v})} \|(({ T}^{-1} \varepsilon - \varphi)+{ T}^{-1}C)^2<
 ({ T}^{-2}\varepsilon- { T}^{-1}\varphi).
\end{split}
\end{equation}

Since the
operator $\Gamma$ of parallel translation preserves the norms of the
vectors, from the last inequality we obtain:
\begin{equation}
\begin{split}
 &\bigg\|\int_{0}^{\tau} \Gamma G_{ S(v(t) + C_{v})} \Big(\frac{d}{dt}S(v(t)+C_v),
 \frac{d}{dt}S(v(t)+C_v) \Big)dt \bigg\|\le\\
 &\int_{0}^{\tau}\Big\|G_{ S(v(t) + C_{v})} \Big(\frac{d}{dt}S(v(t)+C_v),
 \frac{d}{dt}S(v(t)+C_v) \Big)\Big\|dt <\\
 &({ T}^{-1}\varepsilon - \varphi)=k.
\end{split}
\end{equation}

This means that the completely continuous operator $B$ sends the
ball $U_k$ into itself. Hence, by Schauder's principle, $B$ has a
fixed point $v_0(\tau)$ in $U_k$. Then from the definition of
operator $S$ and the usual properties of covariant derivative it
follows that $m_0(\tau) =S(v(\tau)+C_v)$ is a solution of
differential equation \eqref{eq1} (see \cite{yeg,yeg2}). By its
construction the curve $m_0(\tau)$ is a geodesic of the
Levi--Civita connection of Lorentz metric on ${\sf M}$ and, for
it, $m_0(0)=m_0$ and $m_0({ T})=m_1$.

From the equality ${B}v_{0}=v_{0}$ and eq. \eqref{eq4} it follows
that $v_0(0)=0$. Then from the definition of operator $S$ it
follows that $\frac{d}{d\tau}m_0(\tau)_{|\tau=0}=C_{v_0}$, where
the vector $C_{v_0}$ belongs to the light cone of the space ${\sf
T}_{m_0}{\sf M}$ by Lemma \ref{4}. This means that the Lorentz
scalar square of the vector $C_{v_0}$ is negative. Since $C_{v_0}$
is the initial vector of derivative of the geodesic $m_0(\tau)$ of
the Levi--Civita connection on $\sf M$ and since the Lorentz
scalar square of the derivative vector along this geodesic is
constant, the geodesic $m_0(\tau)$ is time-like.
\end{proof}


\begin{thebibliography}{9}

\bibitem{yeg} Gliklikh Yu.E. {\em Global Analysis in Mathematical Physics. Geometric and
Stochastic Methods}. N.Y., Springer-Verlag, 1997. 229 pp.

\bibitem{yeg2} Gliklikh Yu.E. Global and stochastic analysis in the problems of
mathematical physics. Moscow, KomKniga, 2005. 416 pp. (in Russian).


\bibitem{MTW} Misner C., Thorne K., Wheeler J. {\em Gravitation}. N.Y. Freeman,
1973, 1279 pp.


\bibitem{Poltorak1} Poltorak A. On the covariant theory of gravitation. 9th
Interna-tional Conference on General Relativity and Gravitation,
Abstracts.  Jena, 1980. Vol. 2, 516; gr-qc/0403050

\bibitem{Poltorak2} Poltorak A. On the Energy Problem in General Relativity.In:{\em $10$th
International Conference on General Relativity and Gravitation},
Padova, Contributed Papers, Ed.: B. Bertotti, F. de Felice, A.
Pascolini.  1983, Vol.1, 609; gr-qc/0403107

\bibitem{Poltorak3} Poltorak A. Gravity as nonmetricity. General
relativity in metric-afine space $(L_n,g)$; gr-qc/0407060


\bibitem{121} Sachs R.K., Wu H. {\em General Relativity for Mathematicans}.
N.Y.: Springer-Verlag, 1977. 291 pp.

\bibitem{zyk} Zykov P.S. On solvability of two-point boundary value problem for
equations of spray type on Riemannian manifolds. Transactions of
RANS, Series MMMIC. 2004. Vol. 8, No. 1--2, 5--13 (in Russian).

\end{thebibliography}
\end{document}